
\magnification=\magstep1
\baselineskip  22 true   pt
\hsize 5 in\hoffset=.4 true in
\vsize 6.9 in\voffset=.4  true  in
\def\a{\alpha}
\def\b{\beta}
\def\g{\gamma}
\def\d{\delta}
\def\e{\epsilon}

\def\l{\lambda}
\def\om{\omega}
\def\p{\pi}
\def\m{\mu}
\def\n{\nu}
\def\r{\rho}
\def\s{\sigma}

\def\QNE{{F^a_{\mu\n}}}
\def\QNF{{F^a_{\mu 0}}}
\def\QNEB{{{\tilde{F}}^a_{\mu\n}}}
\def\QNEBS{{{\tilde{F}}_{\mu\n}}}

\def\QNK{{1\over{g_5^2}}}

\def\QNII{{{\bar\eta}_{a\mu\n}}}

\def\lnrt{\longrightarrow}
\def\boxpi{{{\partial^2 \Pi}\over \Pi}}
\def\pip{{{\Pi '}\over \Pi}}
\def\prm{\partial_M}
\def\prl{\partial_L}
\def\tilo{{{{\tilde{\om}}_{\m}}}}
\def\xmhat{{{{\hat{x}}_{\m}}}}

\vfil
\headline{\hfill IP/BBSR/92-36.}
\vfil
\centerline {\bf{ Topological Soliton Multiplets in 4+1 Dimensional
 YMCS Theory}}
\vskip 1.0 true cm
\centerline {\bf{C.S.Aulakh${}^{\dagger}$}}
\centerline{ Institute of Physics, Sachivalya Marg}
\centerline{ Bhubaneshwar,751005,India}
\vskip 0.5 true cm
\centerline {and}
\centerline {\bf{V.Soni}}
\centerline{ National Physical Laboratory, Dr. K.Krishnan Marg}
\centerline{ New Delhi,India}
\vskip 1.0 true cm
\centerline{\bf{ ABSTRACT}}

\noindent

\vskip 0.5 true cm
We generalize our results${}^1$ on charged topological solitons (CTS)
in $4+1$ dimensional $SU(3)$ Yang-Mills-Chern-Simons (YMCS) theory to
 $SU(N)$. The $SU(N)$ multiplet structure of two classes of solitons
 associated with the maximal embeddings
 $SU(2)\times U(1)^{N-2}\subset SU(N)$ and $SO(3)\times
U(1)^{N-3}\subset SU(N)$ and the vital role of the $SU(N)$
multiplet of topological currents is
clarified. In the case of the first embedding one obtains a
$^{N}C_{2}-$plet of CTS. In the second, for $N = 3$, one obtains
neutral solitons which, though (classically) spinless, have magnetic moments.
For $N \geq 4$, after modding out the above mentioned non-particulate
feature, one obtains $^{N}C_{3}$ plets of CTS.

\vfil
\footline {{$\dagger$ Email:aulakh\%iopb{@}shakti.ernet.in} {\hskip 2 cm}\hfil
{\hbox{{May 1992}}}}

\eject

\centerline {\bf 1. Introduction}

In a recent paper${}^1$  the existence of charged
topological solitons (CTS) in 4+1 dimensional $SU(3)$
Yang-Mills-Chern-Simons (YMCS) theory was
demonstrated. The relevance of such theories to the ``syncyclonic
scenario"${}^5$ (i.e. of solitonic signals of the possible existence of
extra compact dimensions) and their interest due to connections with
the Skyrmeon model, collective quantization of non-Abelian monopoles
etc is explained in Ref.[1,4,5] and in Section 6.

In this paper we generalize our previous results to the case of
$SU(N)$ and exhibit the central role of a $SU(N)$ multiplet of
topological currents constructed from the CS term in determining the
charge of the solitons. Thus such theories furnish a novel instance where
a topological current couples ``electrically'' (in contrast to the
``magnetic'' coupling of the topological current in the case of 't Hooft-
Polyakov magnetic monopoles) to massless gauge fields.

 Since, in contrast to solitons in YM-Higgs systems,
the symmetry group of the vacuum is unchanged when a
CTS of the sort found in Ref.[1] is present it is natural to expect
that we should obtain a {\it{multiplet}} of solutions with respect to
the unbroken gauge symmetry . We show that in fact these charged solitons
occupy a triplet of the gauge group  $SU(3)$
corresponding to different possible embeddings  $SU(2)\times U(1)
\subset SU(3)$ associated with the CTS solutions.
 This result can be generalized to $SU(N)$ in
which case (generically called Type I) the charged solitons
are associated with embeddings of
$SU(2)\times U(1)^{N-2}\subset SU(N)$ and occupy
${N(N-1)\over 2}$ plets of SU(N).

We then expose a new type (Type II) of {\it{neutral}} soliton
associated with the maximal embedding  $SO(3) \subset SU(3)$
 and with winding number 4 ( cf. the dibaryon embedding of the Skyrmeon).
For a general $SU(N)$ gauge group the relevant embeddings are
 $SO(3)\times U(1)^{N-3} \subset SU(N)$.

 The angular momentum of  both types of solitons  is zero
  due to the orthogonality of the electric and
magnetic fields in the internal space. Nevertheless, the asymptotic
fields are those of a non-abelian magnetic {\it{dipole}}. The presence
of a dipole moment in the absence of an analog of spin is possible due to the
extension of the field configuration i.e it is a mark of the non-particulate
features of the soliton. Solutions related by certain $SU(3)$ rotations
have the same charges but different magnetic moments. Likening
the magnetic moment to an internal degree of freedom one may ``mod it
out'' so as to isolate and highlight the particulate features
of the topological soliton. In the case of Type II, $N=3$ solitons
this merely results in one neutral object.
 For higher N, once one mods out the action of
the $ SU(3)$s relating solutions with identical $SU(N)$ charges but different
magnetic moments one obtains a ${N(N-1)(N-2)\over 6}$-plet of solitons.

This paper is organized as follows.
In Section 2. we discuss the Noether and topological currents of
the YMCS theory and explain their relevance to the charge and
couplings of topological solitons. In Section 3. we review the
analysis of Ref.[1] and simultaneously generalize it to the case $N >
3$.  In Section 4.  we explain how these (Type I)
solitons organize into multiplets of the symmetry group $SU(N)$ of the
vacuum. In Section 5. we consider the case of $SO(3)\times U(1)^{N-3}
\subset SU(N)$ (type II) embeddings and explain its salient features.
In Section 6. we conclude with a discussion of our results along with
their connection with previous and future work.
\vskip 0.7 true cm
\centerline {\bf 2. Currents and Couplings}

We begin with the SU(N) YMCS action on $M^{5}$

$$\eqalign {  S &= \int d^5 x ({1\over {2 g_5^2}} tr {F_{MN}}^2 +
 {{i N_f}\over {48\pi^2}} \omega_5) \cr
 \omega_5 &=\e_{MNLPQ}\quad tr(\prm A_N(\prl A_P A_Q + {3\over 2} A_L A_P A_Q)
 + {3\over 5} A_M A_N A_L A_P A_Q )}
\eqno(1)$$

The field equations are:

$$ D_M F_{MN}^A = -{{N_f g_5^2}\over {128\pi^2}}\e_{NMLPQ}\quad tr(\l^A
F_{ML} F_{PQ}) \eqno(2) $$

Our normalizations and definitions are:

$$\eqalign{ A_M &= A_M^A {{\l^A}\over 2i} \qquad\qquad tr \l^A\l^B =2\d^{AB}\cr
F_{MN} &=F_{MN}^A {{\l^A}\over 2i}=\partial_{[M} A_{N]} + [A_M,A_N]\cr
D_{M} &=\partial_{M}  + [A_M,\quad]\cr
A,B,...&= 1,...,N^2-1 ;\qquad M,N,...= 0,1,2,3,4\cr
\m,\n,...&= 1,2,3,4 ;\qquad a,b,...=1,2,3\cr}\eqno(3)$$

 Our metric convention is $(-++++)$.
Here $ \{\lambda^{A}; A=1.. N^{2}-1\}$  are the Gell-Mann matrices for $SU(N)$
i.e. $\{E_{ij},\tilde E_{ij} \mid 1\leq i < j\leq N\} \bigoplus
\{H_{{k}^{2}-1}, k=2...N\}$ where $\{E, \tilde E\} ,H$ are N x N matrices
spanning the off diagonal set and the Cartan sub-algebra (CSA)
respectively :

$$\eqalign{ [E_{ij}]_{kl} &= \d_{i(k} \d_{l)j}\qquad \qquad
 [ {\tilde E}_{ij}]_{kl} = -i\d_{i[k} \d_{l]j}\qquad \qquad \cr
 H_{k^2-1} &=\l^{k^2-1} = {\sqrt{2 \over{k(k-1)}}} diag(1_{k-1},1-k,0_{N-k})
}\eqno(4)$$

The CTS solutions of the field equations (2) found in Ref.[1] for the
case of $SU(3)$ arose essentially as winding number preserving
deformations of the topological solitons of 5 dimensional YM theory
(i.e instantons of 4 dimensional YM theory reinterpreted
as solitons in one higher dimension)
 due to  the addition of the CS term to the action. Before
presenting the generalisation of our solution to $SU(N)$ in {\hbox{Section 3}}
we first discuss the conserved $SU(N)$ Noether and topological current
multiplets $\{N^{A}_{M}\},\{ T^{A}_{M}\}$.

The Noether currents which arise due to the invariance of the action
(1) under global $SU(N)$ rotations are:

$$\eqalign { N_M^A &=- f^{ABC}
 { {\partial {\cal L}}\over {\partial \partial_MA_N^B}} A_N^C \cr
 &= \QNK f^{ABC}A_N^BF_{NM}^C +{\gamma \over 4} \epsilon_{MNLPQ}
 tr([\l^A,A_N](F_{LP}A_Q+A_QF_{LP} -A_LA_PA_Q))\cr
 &= (N_M^A)^{YM} + (N_M^A)^{CS}}\eqno(5)$$

In addition to $N^{A}_{M}$ there is a multiplet of conserved
topological currents which plays a vital role with respect to the CTS
solutions. It is easily derived as follows. Under an arbitrary
variation of the gauge potential $\d A_{M}$, the CS density
has variation $(\g={N_f\over{48\p^2}})$:

$$\d (i\g\omega_5)=i\g\epsilon_{MNLPQ}
 tr({3\over 4} F_{MN}F_{LP} \d A_Q -
 {1\over 2} \partial_M(F_{NL}A_P+A_PF_{NL} -A_NA_LA_P)\d A_Q) \eqno(6)$$

for a global $SU(N)$ rotation $\d^{A}A_{Q}=[{{\lambda^{A}}\over {2i}},A_{Q}]$ ,
$\d\omega_{5}$ is {\it{identically}} zero.
 On the other hand using $\partial_{M}tr(ABC)=tr D_{M}(ABC...)$
 it is easy to show that for such a variation

$$\eqalign{\d (i\g\omega_5) &=-\partial_MT_M^A=0\cr
T_M^A &= {{3\g}\over 8} \epsilon_{MNLPQ} tr\l^AF_{NL}F_{PQ} -(N_M^A)^{CS}\cr
&=Q_M^A-(N_M^A)^{CS}   }\eqno(7)$$

$\{T^{A}_{M}\}$ is the (identically) conserved multiplet of
topological currents. As indicated the second term in (7 ) is
precisely the negative of the contribution of the CS term to the
Noether curent so that the total $SU(N)$ current is
$J_{M}^{A}=(N_{M}^{A})+ T_{M}^{A}=(N_{M}^{A})^{YM}+ Q_{M}^{A}$.
 The field equations can thus be written in the suggestive forms

$$\partial_MF_{MN}^A=-g_5^2J_N^A\eqno(8a)$$

$$D_MF_{MN}^A=-g_5^2Q_N^A=-g_5^2(T_N^A+(N_N^A)^{CS})\eqno(8b)$$

The form (8a) shows that it is the total current (i.e Noether plus
topological) which acts as a source of the field strength. In
particular the charge of a field configuration will (by definition)
be $\sim \int J_{0}^{A}d^{4}x$. The form (8b) of the field equations
shows that the CS term's contributions to the SU(N) current enter the YM field
equations like an external  current. The coupling of static soliton solutions
$\{{{\hat A}_{M}}\}$ of eqns(8) to weak, static, external potentials
$\{\overline {A_{M}}\}$ clarifies this analogy. To see thus consider
the expression for the field energy to leading order in the external
field. With $A_{M}={\hat A_{M}} +{\overline A_{M}}$ and $\partial_{0}=0$ we
obtain via an integration by parts

$$\eqalign{{\cal E} &=
 {\QNK} \int d^4x({1\over 4} {F_{\m\n}^A}^2 + {1\over 2} {F_{\m0}^A}^2)\cr
 &={\hat {\cal E}} + \int d^4x ({\overline A}_0^A Q_0^A +
 {\overline A}_{\m}^A Q_{\m}^A  -
 2{\overline A}_{\n}^A ({\hat{D}}_{0}{\hat{F}}_{0\n})^A) \cr
 &\qquad + \lim_{R\rightarrow  \infty}\int_{S_R^3} d\Sigma_{\m}
({\overline A}_{\n}^A ({\hat{F}}_{\m\n})^A +
 {\overline A}_{0}^A ({\hat{F}}_{\m 0})^A) + O({\overline A}^2)
  }\eqno(9)$$

where we have used the fact that $\hat A_{M}$ satisfies eqn.(8).
The 4th term in the last line of eqn.(9) is zero for the solutions we
consider. The analogous argument in the absence of the Chern-Simons term would
give zero to leading order in $\{\overline {A_{M}}\}$ even though the
solution in that case has the fields of a dipole. Thus we see that
an external potential couples to solutions of the field equations (8)
couples via the total(topological plus Noether) contribution of the
CS term to the current.

\vskip 0.5 true cm
\centerline {\bf 3. Type I or $SU(2)\times U(1)^{N-2} \subset SU(N)$
Associated Solutions}

We now present the generalization of the analysis of Ref.[1] to the
case $N > 3$ for reference and completeness.
In the absence of the CS term ($N_f=0$) we have the usual self dual
solutions${}^{2,3}$  in the SU(2) Euclidean sector:

$$\eqalign{ A_0&=\partial_0 =0\cr
A_{\m} &=-\QNII {{\l^a}\over 2i} \partial_{\n} ln\Pi =
i{{{\overline\Sigma}_{\m \n}}}\partial_{\n} ln\Pi \cr
\Pi^{-1}\partial^2\Pi &=0 \qquad\qquad \Pi =1
+\sum_{i=1}^K {{\r_i^2}\over {(x-x_i)^2}} \cr
\n &= {1\over{32\p^2}}\int d^4x \QNE \QNEB = K }\eqno(10)$$

where $\r_i,x_i$ are the scale factors and positions of the K solitons and
{\hbox{$\n=K$}}  the total winding number.
 In the above $\{\QNII\},\Pi$
are the anti-self-dual 't Hooft symbols, and ``super potential"
respectively. Fields outside the $SU(2)$ subalgebra  generated by
$\{\lambda^{a}\}$ play no role. The asymptotics are those of a
non-abelian magnetic dipole with zero electric charge (since $F_{\mu
0}=0)$. The field energy is

$${\cal E} = {\QNK} \int d^4x({1\over 4} \QNE^2 + {1\over 2} \QNF^2)
= {{8\p^2\n}\over{g_5^2}}\eqno(11)$$

We now show that these solutions become charged when the effect
 of the C.S. term is included. For simplicity let us work with $\n=K = 1$.
  The field equations are:

$$ D_{M} F_{M\n}^A = {{N_f g_5^2}\over {16\pi^2}}\quad
 tr(\l^A F_{\m 0}\QNEBS) \eqno(12a)$$

$$ D_{\m} F_{\m 0}^A = -{{N_f g_5^2}\over {64\pi^2}}
 tr(\l^A F_{\m \n}\QNEBS) \eqno(12b) $$

It is easy to check that for $A = k^2-1,k=3...N $
the ansatz of (10) gives zero for
the l.h.s of eqn.(12b) while the rhs is proportional to
the Pontryagin density and is hence nonzero.
 Thus (10) is no longer an adequate ansatz when $N_f\neq 0$ and we
are motivated to look for a solution with the modified ansatz:

$$\eqalign{A_{\m}^A &=0\qquad\qquad A\not\in \{1,2,3\}\cr
A_{\m}^a &=-\QNII  \partial_{\n} ln\Pi \qquad\qquad \Pi=\Pi(x^2)\cr
A_0^A &=0=\partial_0 \qquad A\not\in \{k^2-1;k=3..N\}} \eqno(14)$$

Spherical symmetry ensures that the r.h.s of the equation for
${D_{\m}F_{\m 0}^a}$ vanishes as it must for consistency. The only
nontrivial equations are then :
$$\eqalign {{D_{\m}F_{\m\n}^a} &=
-{\a{\sqrt 3}\over 2}{{{\tilde{ F}}_{\m \n}^b}}
\sum_{k=3}^{N}  \partial_{\m} A_0^{k^2-1}
tr(\l^a\l^{k^2-1}\l^b)  \cr
 \partial^2 A_0^{k^2-1} &= {\a{\sqrt 3}\over 8} tr(\l^{k^2-1}
Diag(1_2,0_{N-2}))
 \QNE \QNEB
 =  {\a\over 2}{\sqrt {6\over{k(k-1)}}}\partial_{\mu}{{{\tilde{\om}}_{\m}}}\cr
 \a&={{N_f g_5^2}\over {{32\pi^2}{\sqrt {3}}}}\cr
 {{{\tilde{\om}}_{\m}}}&= \e_{\m\n\l\g}(A_{\n}^a\partial_{\l}A_{\g}^a +
 {{\e_{abc}}\over 3} A_{\n}^aA_{\l}^b A_{\g}^c)}\eqno(15)$$

Thus the Pontryagin density serves as the charge density in a Poisson
equation for the electrostatic potential in the $\{k^2-1,k=3...N\}$ directions.
The solution for $A_0^{k^2-1}$ is simply

$$A_0^{k^2-1}(x) = -{\a\over{16\pi^2}}{\sqrt{6\over{k(k-1)}}}
\int d^4y{1\over{(x-y)^2}} \QNE(y) \QNEB(y)
\eqno(16) $$
and also

$$ \partial_{\m} A_0^{k^2-1} = {\a \over 2}{\sqrt{6\over{k(k-1)}}}
({{{\tilde{\om}}_{\m}}} +
 {{D {{{\hat{x}}_{\m}}}\over {x^3}}}) \eqno(17) $$

The arbitrarinesss represented by D is crucial to the existence of a
charged solution. It is easy to check
(assuming that the deformed superpotential
has the same leading behaviours as $x\lnrt 0$ and $x\lnrt\infty$ as before)
that the leading behaviour of
$\tilo$  as $x\lnrt\infty\quad$  is $O(x^{-7})$ . Thus to obtain the
electric field corresponding to the charge
$Q^{k^2-1}=-{N_f\over{4}} {\sqrt{2\over{k(k-1)}}}\n$
we must choose $D=8$.

The other equation is now

$$ D_{\m} F_{\m\n}^a  =-{{\a^2} \over 2}({\sum_{k=3}^N{6\over{k(k-1)}}})
({{{\tilde{\om}}_{\m}}} +
 {{D {{{\hat{x}}_{\m}}}\over {x^3}}}) {{{ \tilde{F}}_{\m \n}^a}}\eqno(18) $$

Due to spherical symmetry the above equation reduces to

 $$\eqalign{(\boxpi)' -2(\pip)(\boxpi) &=
 {{\a^2}\over 2}({\sum_{k=3}^N{6\over{k(k-1)}}})
 ({{{\tilde\om}}} + {D \over {x^3}}) {\tilde f}\cr
{{{\tilde\om}}} &= -2(\pip)^2 (\pip + {3\over x}) \cr
{\tilde f} &= \pip (\pip +{2\over x})} \eqno(19) $$

To preserve the Pontryagin index and finiteness of energy the
boundary conditions on $\Pi$ are taken to be the same as before i.e.
$\Pi\lnrt {\r^2\over{x^2}}$ as $x\lnrt 0$ and
$\Pi\lnrt 1 +{\r^2\over{x^2}} $ as $x\lnrt\infty$

At first sight the extreme non linearity of these equations seems
intractable. One expects,however, that if one
can solve it in the asymptotic regions $x\lnrt 0,x\lnrt \infty$
to obtain a solution with the same winding number as for $N_f=0$ then
a solution which interpolates between
these regions may be obtained numerically. Evaluating the r.h.s of eqn(19)
in the limit $x\lnrt 0$ one
finds that the leading term is
$-2\a^2(D-8){\sum_{k=3}^N{6\over{k(k-1)}}}/(x^3\r^2)$  while the l.h.s is less
singular. Hence the choice $D =8$ is confirmed. With $D = 8$ one finds
that one can solve for the unknown coefficients in the deformed superpotential

 $$\Pi_0= 1 + {{\r^2}\over {x^2}} + a_1 x^2 + a_2 x^4 + \ldots \eqno(20)$$

to get (${{\a}'}=\a{\sqrt{\sum_{k=3}^N{6\over{k(k-1)}}}}$)

 $$ a_1 =-{{{{\a}'}^2}\over {\r^4}}\qquad\qquad
 a_2 =+{{2{{\a}'}^2}\over {3\r^6}}(1 +{{9{{\a}'}^2}\over {4\r^2}})\eqno(21)$$

and so on. Note that inspite of the nonlinearity of the fermion back
reaction represented by the Chern-Simons term, the deformations are
entirely non singular. Similarly in the $x\rightarrow\infty$ region

  $$\Pi_{\infty}= 1 + {{\r^2}\over {x^2}} +
  {{ b_1}\over{ x^4}} +   {{ b_2}\over{ x^6}}  + \ldots \eqno(22)$$

 $$ b_1 ={{{{\a}'}^2 {\r^2}}\over 3}\qquad\qquad
b_2 ={{{{\a}'}^4 {\r^2}}\over {18}}\eqno(23)$$

A numerical integration of the equation in the intermediate region will
be given elsewhere. The asymptotic behaviour of the electric field is

$$\eqalign{ E_{\m}^{k^2-1} &=-\partial_{\m} A_0^{k^2-1} =
 -{\a \over 2}{\sqrt{{6\over{k(k-1)}}}}
 ({{{\tilde{\om}}_{\m}}} +
 {{D {{{\hat{x}}_{\m}}}\over {x^3}}})\cr
 &={\sqrt{{6\over{k(k-1)}}}} (-{{4{{\a}}}\over {x^3}}{{{\hat{x}}_{\m}}} +
  {{12{{\a}}\r^4}\over{x^7 }} {{{\hat{x}}_{\m}}}) +O(x^{-9})})\eqno(24)$$

The electric charges are

$$Q^{k^2-1}=\QNK\int_{S_{\infty}} d\Sigma_{\m} E_{\m}^{k^2-1} =
-{{N_f}\over{4{}}}{\sqrt{{2\over{k(k-1)}}}}\eqno(25)$$

{\vbox{
The electric field near the origin is

$$E_{\m}^{k^2-1}= -{{12{{\a}}}\over {\r^4}}{\sqrt{{6\over{k(k-1)}}}}
{{x_{\m}}} +O(x^3)\eqno(26)$$
}}

{\vbox{
The magnetic field is

$$\eqalign{\QNE &=-{4\over{\r^2}}(\QNII +
 2  {{{\hat{x}}_{\s}}}{{\bar\eta}_{a\s{[\m}}}{{{\hat{x}}_{\n ]}}})  + O(x^2)
 \qquad\qquad x\lnrt 0\cr
&=-{{4\r^2}\over{x^4}}(\QNII +
 2  {{{\hat{x}}_{\s}}}{{\bar\eta}_{a\s{[\m}}}{{{\hat{x}}_{\n ]}}})  + O(x^{-6})
 \qquad\qquad x\lnrt \infty}
 \eqno(27)$$

The magnetic field is thus asymptotically {\it{dipolar}}.
}}
The CS term does not contribute to the expression for the field
energy :

$$\eqalign{{\cal E} &= {{8\p\n}\over{g_5^2}} +
 {3\over{2 g_5^2}}\int d^4x(\boxpi)^2\cr
&\qquad-{1\over{2g_5^2}}\sum_{k=3}^N
\int d^4x d^4y {\r}_{k^2-1}(x) G_0(x,y){\r}_{k^2-1}(y)\cr
{\r}_{k^2-1}(x) &=-{\a\over 4} {\sqrt{{6\over{k(k-1)}}}}\QNE\QNEB
\qquad\qquad G_0(x,y)=-{1\over{4\pi^2(x-y)^2}}}
\eqno(28)$$

Thus it is the sum of the original uncharged soliton energy plus a
positive definite contribution from the electrostatic energy and
another due to the deviation from self duality. The winding number is
unaffected by the weak deformations exhibited in equations (20)-(23).

$$\eqalign{\n &= {1\over{32\p^2}}\int d^4x \QNE \QNEB =
{1\over{16\p^2}}\int d^4x \partial_{\m}{\tilo}\cr
&={1\over 8} x^3\xmhat\tilo\mid^{\infty}_0=1 }\eqno(29)$$

This can be checked by transforming to a nonsingular gauge using the
usual${}^{3}$ transformation
{\hbox{$U=(x^4+i{\vec{\tau}\cdot{\vec{x}}})/{\mid x\mid}$.}}
The above solutions are parametrized by an arbitrary scale parameter $\r$.
However the energy ${\cal E}$  is not independent of $\r$  .
 Thus one expects that the present solution, if stable at all,
  will relax to that value of $\r$ which minimizes the energy.
  To estimate this value one may
approximate $\Pi$ by $\Pi_{0}$ and $\Pi_{\infty}$ in the regions
$x\in [0,\r],x\in[\r,\infty]$  respectively to obtain

$${\cal E}= {{8\p^2}\over {g_5^2}}+
{{\b g_5^6 N_f^4}\over {\r^4}}
+{{\d g_5^2N_f^2}\over {\r^2}}\eqno(30)$$

Where the last term is the obvious dimensional estimate for the
electrostatic energy in 4 space dimensions and $\b,\d$
 are positive numerical constants.
In the approximation we have used , it appears that our solution is
unstable against growth of the free parameter $\r$ which will tend
to \ $\infty$ so as to saturate the the self-duality lower bound on the energy.
Note however that the corrections to the neutral soliton energy are $O(g_5^2)$
and $O(g_5^6)$ thus it quite possible that quantum corrections may
stabilize $\r$ at a finite value.

Since the electric and magnetic fields are orthogonal in the internal
space, the Poynting vector, and therefore also the usual
 (Belinfante-Bessel-Hagen) expression for the the angular
 momentum of the field configuration, is zero. The nonzero
components of the dipole moment
$\m^a_{\m\n}=4\r^2\QNII\quad $(which can be read off
from eqn.(27)) are a signal of the non-particulate aspects of the soliton.

\centerline{{\bf 4. Multiplet Structure of  Type I CTS}}

As is well known topological soliton solutions of nonlinear field
equations although classical exhibit many of the features of a
{\it{quantum}} point particle. In the present case we wish to show that the
family of solutions obtained by performing a global $SU(N) $ rotation
on the solution given above is analogous to the orbit of states
produced by  $SU(N)$ rotating an ${{N(N-1)}\over 2}$-plet representation
charge eigenstate.

The first element of the analogy is simply the observation that the
basic solution of Section 3.  is analogous to a particle in an $SU(N)$
eigenstate in as much as only the CSA generators $\{Q^{k^{2}-1},k=3...N \}$
are nonzero while the ``raising and lowering" (off diagonal) charges
are zero: in exact analogy with the expectation value of $SU(N)$
generators in a charge eigenstate with  $T_{3}=0$. To determine
what representation of $SU(N)$ our ``state" is a member of, we need to
determine all the distinct ``eigenstates'' (i.e those solutions with
$Q_{A} =0 \quad A \neq k^{2}-1\quad, k = 2...N)$ on the $SU(N)$ orbit
obtained by $SU(N)$ rotating the solution $\{\hat A_{\mu}\}$. Since the field
equations are covariant under such rotations every such rotation gives
a solution which, however, is not in general analogous to an eigenstate
since the off diagonal charge values are nonzero. In fact
if we write $A'_{\mu}=U\hat A_{\mu} U^{\dagger}$ where $U\in SU(N)$
then it is easy to see that $A'_{\mu}$ is also a solution of the
field equations with topological charge $Q' = U Q (\hat A_{\mu})
U^{\dagger}$ ($Q=Q^A{\l^A\over{2i}}$).
 We wish to determine the possible eigenstates reachable via
$U\in SU(N)$. For simplicity consider the case $N = 3$. Writing

$$\eqalign{Q' &=
a \lambda_{3} +{\sqrt 3}b \lambda_{8} ={\sqrt 3}q U \lambda_{8}U^{\dagger}\cr
q &=-{N_f\over{4\sqrt 3}}\cr
U &=\pmatrix{v_{2\times 2} & w_{2\times 1}\cr x_{1\times 2}& y \cr}}\eqno(31)
$$

we wish to find the possible values of (a,b) . Since $det U = 1$ it
follows that

$$(a^2-b^2)b= -q^3 \eqno(32)$$

so

$$a=\pm {\sqrt{{b^3-q^3}\over{b}}}\eqno(33)$$

Further using the unitarity of $U$ and eqn.(31) we have

$$\eqalign{vv^{\dagger} - 2ww^{\dagger} &={1\over q} diag (a+b,-a+b)\cr
vx^{\dagger}-2wy^* &=0\cr
xx^{\dagger}+yy^* &=-{{2b}\over q}}\eqno(34)$$

Since $Q$ is Hermitian $a,b$ are real. From (31) and (34) and
unitarity we have ($w^T=(w_1,w_2)$)

$$\eqalign{w_1w_2^* &=0\cr
\mid w_1\mid{}^2 &= {{q-(a+b)}\over {3q}} \cr
\mid w_2\mid{}^2 &= {{q+(a-b)}\over {3q}} }\eqno(35)$$

Solving eqns.(32)-(35) one finds that the allowed values of $(a,b)$ are
\hfil\break
 $\{ q (0,1), \quad q (\pm{\sqrt 3\over 2},-{1\over 2})\}$.
 So we see that the only charge eigenstates on the orbit
  are those corresponding to a {\bf 3} or ${\bf {\overline 3}}$ of $SU(3)$.

  The literal generalization of this argument to the general case is
algebraically daunting. However there is an alternative more
intuitive way of obtaining the above result which allows one to guess
the result in the general case. In Section {\bf 3} the embedding
of $SU(2)\times U(1)^{N-2}$ in $SU(N)$ was defined by taking the generators
of $SU(2)$ to be $\s_{(1)}^1=\l^1=E_{12},\s_{(1)}^2=\l^2={\tilde{E_{12}}},
\s_{(1)}^3=diag (1,-1,0_{N-2})$ and the commuting $U(1)$s to be generated by
${\s_{(1)}^{k^2-1}= H_{k^2-1}}$ where we have introduced a subscript
(1) to label this (first) embedding. Clearly one can obtain ${}^NC_2-1$ other
 (${{SU(N)}\over {SU(2)\times U(1)^{N-2}}}$ related) ``placements'' by
 interchanging the role of the pair of N-plet indices (12) with some other
 unequal ordered pair ($i<j$) ( (13),(23) for SU(3) etc.). Then $\s_3$ will
 have 1 and -1 in the i'th and j'th diagonal place and zero elsewhere,
 $\s_8$ will have ${\sqrt {1\over 3}}$ in the i'th and j'th diagonal
 place etc.The solutions corresponding to these options  are
$A^{(i)}_{\mu}=\hat A_{\mu}^{a}{\sigma_{(i)}^a\over 2i}$
It is easy to calculate the topological
  charges for each placement (with reference to the canonical
Gell-Mann basis i.e by expanding the $\s_{(i)}^A$ for each placement
in terms of the $\{\l^A\}$ ) and confirm that they correspond
precisely to the weights of a ${}^NC_2$-plet of $SU(N)$. Of course
the anti-solitons occupy the conjugate representations.

\vskip 0.5 true cm
{\centerline {\bf 5.Type II or $SO(3)\times U(1)^{N-3} \subset SU(N)$
Associated Solutions.}}

As we have shown in Section {\bf 4}, when the generators $\{ \sigma^{a};
a = 1,2,3\}$ in terms of which the ansatz for $A_{\mu}$ is formulated
receive an $SU(2)\times U(1)^{N-2}$ embedding in $SU(N)$ then any
non-trivial solution found under the ansatz is necessarily charged.
We now show that if the $\{ \sigma^{a}\}$ are embedded as {\it{just}}
 $ SU(2) (SO(3)) $in SU(3) then a neutral solution exists. To see this note
that in the Gellmann-basis the structure constants are:

$$\eqalign{f_{123}&=2 f_{345}=-2 f_{367}= {2\over{\sqrt {3}}}f_{458}=
{2\over{\sqrt {3}}}f_{678}=1\cr
f_{147}&= f_{516}= f_{246}= f_{257}={1\over 2}\cr}\eqno(36)$$

\noindent The first group provided us with the 3 basic placements of
$SU(2)\times U(1)$ . Similarly the second group provides us with the 4
basic placements of the $SO(3)\subset SU(3)$ maximal embeddings i.e
$\{\s^a\}\equiv\{\s^1=2\l^1,\s^2=2\l^4,\s^3=2\l^7\}
\equiv \{2\l^{\overline a},{\overline a}=1,4,7\}$ etc.
Now $F_{\mu\nu}^{\overline a}=\partial_{[\mu}A^{\overline a}_{\nu ]} +
{1\over 2} \epsilon^{{\overline {abc}}}
A_{\mu}^{{\overline b}}A^{\overline c}_{\nu}$
so if we take
$A_{\mu}^{{\overline a}}=-2\QNII \partial_{\n}ln \Pi(x^2)
\equiv 2 { {{\hat A}_{\mu}^{\overline a}}}$
spherical symmetry gives
$F_{\mu\nu}^{{\overline a}}\tilde F^{{\overline b}}_{\mu\nu}\sim
\delta^{{\overline {ab}}}$
 so that the r.h.s. of the time component of the field equation
(8) is proportional to
$tr (\lambda^{A}\lambda^{\overline b}
\lambda^{\overline b})\sim tr\lambda^{A}=0$.
Thus  we can set $A_{0}=0$ and then the field equations
reduce to $({\hat D}_{\m}{\hat F}_{\m\n})^{\overline a}=0$  where
${{\hat{D}}_{\m}}^{{\overline {ac}}}=
\d^{{\overline {ac}}} \partial_{\m} +  \epsilon^{{\overline {abc}}}
{\hat A}_{\mu}^{{\overline b}}$  and these are
solved by the usual self dual form of $\Pi$. Note however that since
$F_{\mu\nu}^{{\overline a}}=2\hat F^{\overline a}_{\mu\nu}$
the winding number is now 4!! (recall the
``dibaryon" embedding of the Skyrmeon in
$SU(3)$). Thus we have obtained an electrically neutral soliton with
no analog of  spin at the classical level but a nonzero (non-abelian) magnetic
dipole moment. Note that after collective quantization of the global
gauge zero modes these solitons can pick up${}^5$ a charge and spin {\it{a la}}
the Skyrmeon flavor charges and spin${}^9$.

 When we consider $SO(3)\times U(1)^{N-3}$ embeddings
in $SU(N), N\geq 4$ the solitons once more become charged. There are
${}^NC_3$ sets-of-4 ${{SU(N)}\over {SO(3)\times U(1)^{N-3}}}$
related embeddings analogous to the
single set-of-4: $\{ (147), (516), (246), (257)\}$ for $SU(3)$. This
becomes obvious when we write the above example as \hfil\break
$\{ (E_{12},E_{13},{\tilde E}_{23}),({\tilde E}_{13},E_{12}, E_{23}),
( {\tilde E}_{12},E_{13},E_{23}),
({\tilde E}_{12},{\tilde E}_{13},{\tilde E}_{23})\}$. \hfil\break
As explained
above in the case of $SU(3)$  the topological charges for each of these
4 choices are zero since $tr(\l^A \l^{\overline b}\l^{\overline b}) \sim
 tr (\l^A diag(1,1,1))$. For $N\geq 4$, however, one gets in place of $1_3$
 an $N\times N$ diagonal matrix with three diagonal entries equal to
1 and zero elsewhere. This matrix does not have zero trace with $\{\l^{k^2-1},
k=4...N\}$ and hence {\it {all four}} of above embeddings have the
{\it same}  topological charges. Similarly one may chose ${}^NC_3-1$
other sets of 3 unequal indices to fill
out the set of weights of a ${}^NC_3$ of
$SU(N)$.  If we mod out
the action of the (different)  $SU(3)$s which relate these sets-of-4
then one obtains a multiplet with the quantum numbers (topological
charges) of a ${}^NC_3$-plet of $SU(N)$.

The analogy with a point particle becomes some what obscure in the
present case. In the case of $SU(3)$ the $SO(3) \subset SU(3)$ embedding yields
a soliton with zero $SU(3)$ topological charge. The angular momentum of
the field configuration is zero (since the electric field and
therefore the Poynting vector are zero). Thus we have a ``spinless"
neutral lump with a (non-Abelian) magnetic dipole moment. Clearly the
presence of the dipole moment is a mark of the extension of this
object. The modding out of the action of SU(3) may be thought of as a
``compression" to remove the non-particulate features (ie its magnetic
moment) of the lump from view.
\vskip .5 true cm

\centerline{\bf{6. Discussion}}

In Ref[4] we suggested that an
examination of the properties of a certain class of ``co-winding''
solitons (``syncyclons") generically present in higher dimensional
field theories (i.e defined on space-time extended beyond 4
dimensions by a compact space) is called for. Such solitons would
appear as point, string or sheet vacuum defects to our three
dimensional low energy eyes and might thus signal the presence of
otherwise unobservable extra dimensions.
The simplest paradigmatic example of such solitons is furnished by
Yang-Mills (YM) theory on $M^{4}\times S^{1}$. The basic solution is
nothing but the periodic instanton or ${\it caloron}^6$
reinterpreted as a soliton on $M^{4}\times S^{1}$. Classically, such
solitons have no (non-abelian) electric charge and, in the
Kaluza-Klein (KK) limit $r>>R$ (where R is the radius of $S^{1}$ and
$r$ the 3-dimensional distance),
their fields are those of a non-abelian magnetic dipole together
with a scalar potential coming from the extra compoment of the
vector potential${}^{7,4}$.

In order to sketch a believable
picture of the properties of the above generic class of solitons
it is necessary to take into account the
effects of coupling to fermions and the quantization of global gauge
collective coordinates since these can radically change the quantum
numbers labelling the soliton states due to fermion number
fractionalization${}^8$ and zero point (quantum) rotation in the
internal non-abelian  space. In our particular example, using standard
results on fermion number fractionalization in odd dimensions${}^8$,
one finds that the soliton picks up an abelian charge
{\hbox {$-{{\nu q}\over 4}$}}
 ($\nu$ the Pontryagin index) from each fermion of abelian charge
q (Baryon/Lepton number, Electric charge. etc).
Similarly non-Abelian charge fractionalization occurs and the term
 in the gauge effective action which represents
this effect is nothing but the 5 dimensional Chern Simons term times
$N_{f}$. This term is an essential part of the low-momentum
 approximation of the fermion
determinant in the gauge field background. As we have seen, inclusion
of this term in the bosonic effective action yields charged
topological solitons (non abelian topological ``electropoles'').
 In order to fully determine
the gauge quantum numbers of these soliton one must quantize the
collective coordinates associated with global rotations in the group space.
This problem bears a close analogy to that of determining
the quantum numbers of the Skyrmeon${}^7$. The roles of flavor
and color are interchanged and
the Chern-Simons term bears analogies to the WZW term which also arises
by integrating out fermions coupled to the Skyrme field . However it
appears${}^5$
that in contrast to the Skyrme case this term does {\it not} modify
the Collective Quantization constraint equation which arises ${}^7$.
Rather it only contributes a topological charge  in addition to the
charges that arise due to quantization${}^5$.    The
significant generalization is that the present system is gauge
invariant while the Skyrme-WZW Lagrangian has only a global (flavour)
invariance. In the Skyrmeon case the
effect of the WZW term in the field equations is to produce a
deformation of the original soliton without obviously changing its
external interactions (as opposed to its flavour and spin quantum
numbers which obtain a non-trivial modification due to the
contribution of the WZW term to the above mentioned
Collective Quantization  constraint).
 On the contrary, in the present case, the as we have shown
the deformed solution carries  (topological) $SU(N)$ electric charges
 which act as sources of electric fields and couple to external fields.
In the light of the above, we expect that the low-lying soliton states
may be represented as outer products of $SU(N)$ topological charge
states and $SU(N)$ eigenstates coming from the Collective
Quantization proceedure${}^5$

 Charged solitons arise in YMCS-Higgs systems in 2+1
dimensions as well${}^{10}$ where the presence of the CS term allows
charged vortex solutions to exist.
In 2+1 dimensions the CS term makes the gauge particles massive so that
the induced charge couples to a short range force carrier. The
integral of the Higgs field charge density is proportional to the
quantized flux (divided by 4) so that there too the electric charge
is fractional and topological in nature.
 The existence of charged vortex solutions
in 2+1 dimensions can be seen as the analog of the ``Witten effect''
in 4 dimensions whereby the addition of a CP violating $\theta$ term
(i.e a Pontryagin density) to the YM-Higgs lagrangian results in
charged monopole solutions${}^{11}$.  One therefore expects that
such a conversion of ``magnetic" charge neutral solitons to charged
ones occurs in all odd dimensions provided the theory without the CS
term has a topological soliton solution. Note however that in the present
case the magnetic fields are asymptotically {\it{dipolar}}. Since the
magnetic fields are dipolar they do {\it not } (in contrast to the
situation for a single ``grand unified '' monopole ${}^{12}$) obstruct the
organization of the charged excitations that arise on quantization of
the global gauge collective coordinates into multiplets of the
symmetry group of the vacuum${}^5$ ; thus providing a neat
illustration of the ``dipole evasion'' discussed in references [12].

As a preliminary to the quantization${}^5$
of the collective coordinates of solitons in YM theory on
$M^{4}\times S^{1}$, and also because of their intrinsic interest
we have studied static solutions of the YMCS system in $M^{5}$. The
generalization of our results to $M^{4}\times S^{1}$ and the
situation in higher odd dimensions will be reported separately.
 We have shown that the YMCS theory in
4+1 dimensions like its be-Higgsed cousins in 2+1 dimensions exhibits
fascinating solitonic behaviours. The most remarkable and novel
of these is the way in which a non-abelian multiplet structure
emerges naturally, with
the topological current multiplet associated with the CS term playing
a vital role and furnishing a novel instance  of a topological
charge coupling ``electrically'' to  massless gauge fields.
 The present [1,4,5] papers are
but a beginning in unravelling the intricacies of their behaviour.

\vskip 1.0 true cm

\noindent {\bf Acknowledgements:} We are grateful to P.Panigrahi
for  useful discussions.

\vfil
\eject

\centerline {\bf{References }}

\item {1)} C.S.Aulakh , {\it{ Charged Topological Solitons
in 4+1 Dimensional YMCS Theory}}, IP/BBSR/92-27, to be published in
Mod. Phys. Lett. {\bf{A}}.
\item {2)} G.'t Hooft (unpublished); Phys. Rev. D14 (1976)3432.
\item {} E.Corrigan and D.Fairlie, Phys. Lett. B67 (1977)69.
\item {} R.Jackiw, C.Nohl. and C.Rebbi, Phys. Rev. D15 (1977)1642.
\item {3)} R.Rajaraman, {\it Solitons and Instantons}, North Holland,
1982.
\item {4)} C.S.Aulakh, {\it{Syncyclons or Solitonic Signals from Extra
Dimensions}},
\item{} IP/BBSR/92-14, to be published in Mod. Phys. Lett {\bf{A}}.
\item {} See also A.Strominger Nucl. Phys. B 343 (1990) 167,
M.Kobayashi, Prog. Theor. Phys. 74(1985)1139.
\item {5)} C.S.Aulakh and V.Soni, {\it{Collective Quantization of Topological
Solitons in 4+1 Dimensional YMCS Theory}}, IP/BBSR/92-37, to appear.
\item {6)} B.J.Harrington and H.K.Shepard, Phys. Rev. D17 (1978) 2122.
\item {7)} D.Gross, R.Pisarski, L.Yaffe, Rev. Mod. Phys. 53 (1981) 43.
\item {8)} A.Niemi and G.W.Semenoff, Physics Reports 135 (1986) 99
and references therein ;Phys. Rev. Lett. 51 (1983) 2077.
\item {9)} E.Witten, Nucl. Phys. B223 (1983) 422.
\item{} A.P.Balachandran, V.P.Nair, S.G.Rajeev, and A.Stern, Phys.
Rev. Lett. 49,(1982)1124; 50(1983)1630(E).
\item{} Phys. Rev. D27(1983)1153;D27(1983)2772(E).
\item{} S.Jain and S.R.Wadia , Nucl. Phys. B258(1985) 713.
\item {10)} S.K.Paul and A.Khare, Phys. Lett. B174 (1986) 420, B182
(1986) 414.
\item {} C.N.Kumar and A.Khare ,Phys.Lett. B 178 (1986) 385.
\item{} H.J.De Vega and F.A.Schaposhnik ,Phys. Rev. Lett. 56 (1986) 2564.
\item {11)} E.Witten,Phys.Lett. B86 (1979) 283.
\item {12)} A.Abouelsaood, Nucl. Phys. B226 (1983) 309.
\item {} P.Nelson, Phys. Rev. Lett. 50 (1983) 939.
\item {} C.P.Dokos and T.N.Tomaras, Phys. Rev. D 21 (1980) 2940.
\item {} S.Coleman and P.Nelson, Nucl. Phys. B237 (1984)1.
\vfil
\eject
\end